\documentclass[conference]{IEEEtran}
\usepackage{amsmath}
\usepackage{amsfonts}
\usepackage{amstext}
\usepackage{amssymb}
\usepackage{mathtools}

\usepackage[hyphens]{url}
\usepackage[hidelinks]{hyperref}
\hypersetup{breaklinks=true}
\usepackage{cite}

\usepackage{enumerate}
\usepackage{caption}
\usepackage{subcaption}

\usepackage{algorithm}
\usepackage{algpseudocode}
\usepackage[english]{babel}

\usepackage{epsfig}
\usepackage{xcolor}
\usepackage{color}
\usepackage{graphicx}
\usepackage{lipsum}
\usepackage{pgfplots}
\pgfplotsset{compat=newest}
\usepackage{tikz}
\usetikzlibrary{plotmarks}
\usetikzlibrary{arrows.meta}
\usetikzlibrary{shapes,arrows}
\usetikzlibrary{intersections}
\usepgfplotslibrary{patchplots}

\newcommand{\R}{\mathrm{R}}

\newcommand{\BS}{\mathrm{BS}}
\newcommand{\UE}{\mathrm{UE}}

\DeclareMathOperator*{\argmax}{arg\:max}

\addto\captionsenglish{\renewcommand{\figurename}{Fig.}}

\begin{document}

\title{Reinforcement Learning-based Joint Handover and Beam Tracking in Millimeter-wave Networks}
\author{\IEEEauthorblockN{Sara Khosravi\IEEEauthorrefmark{1},
Hossein S. Ghadikolaei\IEEEauthorrefmark{3}, Jens Zander\IEEEauthorrefmark{1}, and Marina  Petrova \IEEEauthorrefmark{1}\IEEEauthorrefmark{2}}\\
\IEEEauthorblockA{ \IEEEauthorrefmark{1}School of EECS, KTH Royal Institute of the Technology, Stockholm, Sweden,\\   \IEEEauthorrefmark{2} Mobile Communications and Computing, RWTH Aachen University, Germany, \IEEEauthorrefmark{3} Ericsson Research, Sweden\\
Email: \{sarakhos, jenz, petrovam\} @kth.se, hossein.shokri.ghadikolaei@ericsson.com
}}

\maketitle
\begin{abstract}
In this paper, we develop an algorithm for joint handover and beam tracking in millimeter-wave (mmWave) networks. The aim is to provide a reliable connection in terms of the achieved throughput along the trajectory of the mobile user while preventing frequent handovers. We model the association problem as an optimization problem and propose a reinforcement learning-based solution. Our approach learns whether and when beam tracking and handover should be performed and chooses the target base stations. In the case of beam tracking, we propose a tracking algorithm based on measuring a small spatial neighbourhood of the optimal beams in the previous time slot. Simulation results in an outdoor environment show the superior performance of our proposed solution in achievable throughput and the number of handovers needed in comparison to a multi-connectivity baseline and a learning-based handover baseline.
\end{abstract}


\section{Introduction}
Millimeter-wave (mmWave) is a key radio access technology for beyond 5G communication systems, offering ultra-high data rates due to a large amount of free spectrum \cite{rappaport2013millimeter}. However, due to the fewer scattering paths and significant penetration loss, mmWave links are vulnerable to static or dynamic obstacles. To overcome such severe loss, both base station (BS) and user equipment (UE) may need directional communication using a large number of antennas, which may result in frequent misalignment of beams due to mobility and blockage. 
Hence, finding and maintaining the optimal beam directions (beam alignment) is necessary. The lengthy period to achieve the beam alignment (hundreds of milliseconds to seconds \cite{Hassanieh}) results in a high cell search time or BS discovery time in mmWave systems. As reported in \cite{3gppRadioResource}, the BS discovery time which is the time required to search the target BS when the handover command is received by the UE is about $200$ ms. Moreover, to improve the capacity and coverage the density of the BSs is usually high in mmWave systems \cite{rappaport2013millimeter}. Hence, conventional handover methods based on instantaneous received signal power can cause unnecessarily frequent handovers and a ping-pong effect. This leads to a severe drop in service reliability. Therefore, fast BS discovery (finding target BS in the handover process), and efficient handover execution techniques, will be required to use the full promise of mmWave cellular networks. 

The spatial mmWave channel can be approximated by a few dominant paths, where each path can be defined with its angle of departure (AoD), angle of arrival (AoA) and gain \cite{heath2016overview}. Hence, one can only estimate these path parameters instead of a large dimensional channel matrix \cite{sun2019beam, zhang2022training}. The process of identifying the dominant paths is called beam training.  However, due to the dynamic environment, frequent beam training may cause high overhead\footnote{Overhead depends on the training time compared with the changes in the environment.}. Temporal correlation of spatial mmWave channel can be employed to accelerate the beam training process by tracking the variation of the dominant path directions \cite{zhang2022training}. 
\subsection{Related Work}
To address the link failure and throughput degradation in a dynamic environment, the multi-connectivity technique has been vastly analyzed  in literature \cite{ozkocc2021impact,multicon3}. In this technique, the UE keeps its connection to multiple BSs (either at mmWave band or sub-6 GHz band). However, power consumption, synchronization and the need for frequent tracking are the main challenges.
In the 3GPP standard (release 16) two handover techniques are introduced to improve the link robustness during mobility: dual active protocol stack (DAPS), and conditional handover (CHO) \cite{3GPPMulticonnectivity}. In the DAPS, the connection to the current serving BS is maintained until the connection to the target BS is fully established. In the CHO, the UE is configured with multiple target BSs. During the handover, the UE can select one of the configured BSs as the target BS during the RRC reconfiguration message. Although CHO can decrease the handover failure probability, it may increase the handover latency if the UE asks for multiple handovers during a single RRC reconfiguration \cite{ozkocc2021impact}.

Applying machine learning as the main decision-maker tool to make the optimal handover decision and choose the target BS has been also studied in the literature \cite{learninhRL,saraTCCN}. The authors in \cite{learninhRL} proposed a reinforcement learning (RL) based handover policy to reduce the number of handovers while keeping the quality of service in heterogeneous networks. In \cite{saraTCCN} an intelligent handover method based on choosing the backup solution for each serving link to maximize the aggregate rate along a trajectory has been proposed.  

In terms of beam tracking, authors in \cite{patra2015smart} applied the correlation of spatial mmWave channel in adjacent locations and proposed the beam steering method based on searching over a small angular space in the vicinity of the previously known valid beams. The authors in \cite{zhang2022training} applied machine learning to the tracking procedure to extract useful information from the history of AoD tracking. 

All the aforementioned works only take handover or beam tracking issues into account. Additionally, they do not study the impact of selecting beam tracking and handover on the achieved throughput of the UE along its trajectory and instead focus on the achieved rate as the primary performance metric.
\subsection{Our Contributions}
In this paper, we develop a novel joint handover and beam tracking algorithm in a mmWave network under mobility. The algorithm aims to associate the UEs to BSs that maximize the sum achieved throughput along the trajectory and ensure the achieved throughput in each location of the trajectory is higher than a pre-defined threshold. 
The user association process is defined as the process of determining whether a user is associated with a particular BS before data transmissions commence. In the case of handover, the UE is associated with a new BS, whereas in the case of beam tracking, the UE remains associated with the serving BS from the previous time slot.
 The main contributions of our paper are summarized as below:
\begin{itemize}
    \item \textit{System Modeling}: We model the user association problem as a non-convex optimization problem. Unlike the existing works in the literature, we consider achieved throughput as the main performance metric to measure the effect of handover or beam tracking on the UEs' quality of service.
    \item \textit{Learning-based Solution}:  The objective function in our proposed user association problem highly depends on the user association mechanism. We utilize the reinforcement learning (RL) algorithm to approximate the solution to this problem. The aim is to decide whether to run a beam tracking algorithm or a handover algorithm. 
    
    \item \textit{Joint Handover and Beam Tracking Algorithm}: In the case of a handover decision, the target BS will be recognized as the output of the RL algorithm. In the case of beam tracking, the search space will be defined based on our proposed tracking algorithm by searching the directions in the small spatial neighbourhood of the previously selected optimal directions. 
    
     \item \textit{Empirical Evaluation}: We apply ray tracing with a real building data map as the input. The results show the effectiveness of our proposed method in achieving throughput along trajectories and decreasing the number of handovers.
     
\end{itemize}

The rest of the paper is organized as follows. We introduce the system model and problem formulation 
in Section \ref{system model}. In Section \ref{BF}, we propose our method. We present the numerical results in Section \ref{results} and, conclude our work in Section \ref{conclusions}.

\textit{Notations:} Throughout the paper, vectors and scalars are shown by bold lower-case ($\mathbf{x}$) and non-bold ($x$) letters, respectively. The conjugate transpose of a vector $\mathbf{x}$ is represented by $\mathbf{x}^H$. We define set $[M]:=\{1,2,..,M\}$ for any integer $M$. The indicator function $1\{\cdot\}$ equals to one if the constraint inside $\{\cdot\}$ is satisfied.
\section{System Model and Problem Formulation}\label{system model}
In this section, first, we introduce the mmWave channel model. Then, we present the user association problem formulation.

We consider a downlink communication with $| \mathcal{B}|$ mmWave BSs, where each is equipped with $N_{\BS}$ antennas, communicating with a single antenna mobile UE. 
We consider analog beamforming with a single RF chain. We assume all BSs allocate equal resources to their serving UEs.
The channel between BS $j \in \mathcal{B}$ and its serving UE during time slot $i$ is \cite{akdeniz2014millimeter}:
\begin{equation}
 \mathbf{h}_j=\sum_{\ell=1}^{L} h_{\ell}\mathbf{a}^{H}(\phi_{\ell},\theta_{\ell}),
\end{equation}
where $L$ is the number of available paths. Each path $\ell$ has complex gain $h_{\ell}$ (include path-loss) and horizontal $\phi_{\ell}$ and vertical $\theta_{\ell}$, AoD. Due to the notation simplicity, we drop the index $j$ and $i$ from the channel parameters. 
The array response vector is $\mathbf{a}(.)$ where its exact expression depends on the array geometry and possible hardware impairments. The signal-to-noise ratio (SNR) in time slot $i$ is 
\begin{equation}
 \text{SNR}^{(i)}_j= \frac  {p \lvert\mathbf{h}_j^H\mathbf{f}_j\rvert^2}{\sigma^{2}},
 \label{SNR}
\end{equation}
where $\sigma^2$ is the noise power, $p$ is the transmit power, $\mathbf{f}_j\in\mathcal{C}^{N_{\text{BS}}}$ is the beamforming vector of BS $j$.

We define variable $x^{(i)}_j\in\{0,1\}$ for $j\in\mathcal{B}$ as an association indicator in time slot $i$, where is equal $1$ if UE is associated to the BS $j$ and $0$ otherwise.
Hence, the achieved rate per second per hertz in time slot $i$ is
\begin{equation*}
\R^{(i)} = x_{j_S}^{(i)}\log_2(1+\text{SNR}^{(i)}_{j_S})
= \sum_{j\in\mathcal{B}}x^{(i)}_j\log_2(1+\text{SNR}^{(i)}_j),
\end{equation*}
where $j_S$ is the index of the serving BS of the UE during time slot $i$. Here, we assume each UE is served by only one BS. 

We define the achievable throughput per hertz of the UE by multiplying its rate by the data transmission time as
	
	\begin{equation}
\Gamma^{(i)}
=(1-\frac{\tau^{(i)}_b}{\tau_c})\R^{(i)},
\label{throughput}
\end{equation}
where, $\tau^{(i)}_b$ is the beam training duration which may have a different value in each time slot $i$, and $\tau_c$ is the duration of the time slot that is a fixed value for all time slots, see Fig. \ref{fig:beam training}.

\subsection{Beam Training and Beam Tracking}
As depicted in Fig. \ref{beam training a}, when the UE is connected to a BS $j\in \mathcal{B}$, initial beam training is performed by sending pilots over all combination of the beam directions in the codebook during $\tau_b$. Based. on the UE's feedback of the received signal strength (or estimated SNR), the best beam pair directions are selected. Then, the BS and the UE would use this direction ($\phi_{\ell^\star},\theta_{\ell^\star}$) during the data transmission phase. The beamforming vector, $\mathbf{f}$ is chosen to maximize the achievable rate of the UE. Due to the monotonicity of the logarithm function, this is equivalent to maximising the SNR term in \eqref{SNR}. Hence 



	\begin{equation}
	\begin{aligned}
	& \underset{\mathbf{f}_j\in \mathcal{F}}{\mathbf{f}_j^*=\argmax}
	& |\mathbf{h}_j^H\mathbf{f}_j|^2 
	\end{aligned}
		\label{codebook}
	\end{equation}
where $\mathcal{F}$ is the beamforming codebook that contains all the feasible  beamforming vectors. The n-th element of the codebook $\mathcal{F}$ is defined as
$\mathbf{f}(n)=\mathbf{a}(\phi_n, \theta_n)$, where $(\phi_n, \theta_n)$ are steering angles and $\mathbf{a}(.)$ is the array response vector. 

When the BS continues serving the same UE in a consecutive time slot, only searching the neighbouring beam directions of the main directions can be sufficient to maintain the link quality. This process is called beam tracking. As shown in \mbox{Fig. \ref{bema training b}}, the duration of $\tau_b$ is
much smaller than the initial beam training duration.

\tikzset{%
	bodyy/.style={inner sep=0pt,outer sep=0pt,shape=rectangle,draw,thick},
	dimen/.style={<->,>=latex,thin,every rectangle node/.style={fill=white,midway,font=\sffamily}},
	dimen1/.style={->,>=latex,thin,every rectangle node/.style={fill=white,midway,font=\sffamily}},
	symmetry/.style={dashed,thin},
}
\begin{figure}[t]
	\centering
	
	\begin{subfigure}{.5\textwidth}
  \centering
  \scalebox{0.7}{\hspace{-3mm}\small{\begin{tikzpicture}

    
    \node [bodyy,thick,minimum height=1cm,minimum width=28mm,anchor=south west, fill=blue!15!white!] (S) at (0.2,0) {Initial beam training};

\node [bodyy, thick, minimum height=1cm,minimum width=7cm,anchor=south west] at (3,0) {Data Transmission};

\draw[densely dotted,thick] (0.2,1) -- ++(0,+0.4) coordinate (D1) -- +(0,+3pt);
\draw[densely dotted,thick] (3,1) -- ++(0,0.4) coordinate (D2) -- +(0,3pt);
\draw [dimen,blue] (D1) -- (D2) node {$\tau_b$};

\draw[draw=none] (0.2,1.5) -- ++(0,+0.4) coordinate (D11) -- +(0,+5pt);
\draw[draw=none] (9.9,1.5) -- ++(0,0.4) coordinate (D3) -- +(0,7pt);
\draw [dimen,blue] (D11) -- (D3) node {$\tau_c$};


\end{tikzpicture} }}
  \caption{}
  \label{beam training a}
\end{subfigure}
\begin{subfigure}{.5\textwidth}
  \centering
 \scalebox{0.7}{\hspace{-3mm}\small{\begin{tikzpicture}

\node [bodyy,thick,minimum height=1cm,minimum width=12mm,anchor=south west, fill=blue!15!white!] (S) at (0.2,0) {Tracking};

\node [bodyy, thick, minimum height=1cm,minimum width=8.5cm,anchor=south west] at (1.4,0) {Data Transmission};

\draw[densely dotted,thick] (0.2,1) -- ++(0,+0.4) coordinate (D1) -- +(0,+3pt);
\draw[densely dotted,thick] (1.4,1) -- ++(0,0.4) coordinate (D2) -- +(0,3pt);
\draw [dimen,blue] (D1) -- (D2) node {$\tau_b$};


\end{tikzpicture} }}
  \caption{}
  \label{bema training b}
\end{subfigure}
\caption{$\tau_c$ is the time slot duration. $\tau_b$ is (a) the initial beam training duration when the UE is associated with the new BS (handover case), (b) the beam tacking duration when the serving BS is the same for the consecutive slots. }
\label{fig:beam training}
\end{figure}
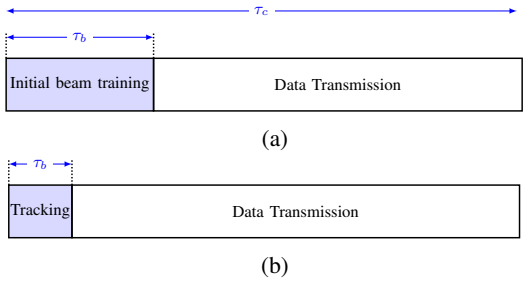

\subsection{Problem Formulation}
The UE association depends on the channel quality between the BS and the UE. Due to UE mobility or temporary blockage, the channel quality changes and consequently the UE association. Based on the UEs' velocity, we determine how quickly the channel quality can change and predict the time at which the current UE association needs to be updated. We define $T_{\text{A}}$ seconds as the frequency of updating the association. Hence, we need to make the decision every $T_{\text{A}}$ whether to run the handover execution or beam tracking procedure if $\text{SNR}$ is lower than the pre-defined SNR threshold ($\text{SNR}_{\text{thr}}$). Note that we can have an on-demand reactive handover at any time slot if the link toward the serving BS fails abruptly. However, with a proper choice of $T_A$, the frequency of those reactive events could be very small. 
We define the duration of the trajectory as $M$ and consider the discrete time index $i$ to describe the association update at each interval.

The goal is to maximize the aggregate throughput of the UE along the trajectory while ensuring the achieved throughput in each time slot $i$ is higher than a predefined threshold. To this end, we define functions $F_1$ and  $F_2$ as 

\begin{itemize}
\item $F_1$ is the averaged throughput along the trajectory as
\begin{IEEEeqnarray*}{rCl}
F_1
&=&\sum_{i=1}^M \mathbb{E}\left[\Gamma^{(i)}\right],
\end{IEEEeqnarray*}
where the expectation is with respect to the randomness of channel fading and the blockage, $M$ is the duration of the trajectory, and $\Gamma^{(i)}$ is defined in \eqref{throughput}.

\item $F_2$ is the expected number of time slots whose throughput is lower than the threshold ($\Gamma_{\rm thr}$).
\begin{IEEEeqnarray*}{l}
F_2
= \mathbb{E}\left[\sum_{i=1}^M 1\left\{ \Gamma^{(i)}\leq \Gamma_{\rm thr} \right\}\right] 
= \sum_{i=1}^M \Pr\Big\{
 \Gamma^{(i)} \leq \Gamma_{\rm thr} \Big\}.
\end{IEEEeqnarray*}

\end{itemize}

We formulate the user association at time slot $i \in {[M]}$ as an optimization problem which involves finding the $x^{(i)}_j$ corresponding to the association indicator as 
\begin{subequations}
\begin{alignat}{3}
&\max_{\substack{\{x^{(i)}_j\}_{i,j} 
}}
&\quad& F_1- \lambda F_2 \label{eq:optProb} \\
&\qquad\quad\mathrm{s.t.}
&& \sum_{j\in \mathcal{B}}x^{(i)}_j= 1, \forall , i\in[M] \label{eq:constraint1}\\
&&& x^{(i)}_j\in \{0,1\}, \quad\forall  j\in \mathcal{B}, i\in[M] \label{eq:constraint4}
\end{alignat} \label{optimization}
\end{subequations}
\!\!\!where $\lambda$ is a large constant controlling the importance of $F_2$. Constraint \eqref{eq:constraint1} guarantees that each UE is served by one BS. 

The optimization problem \eqref{optimization} is nonlinear. Solving this optimization problem requires 
estimating the expectation value in $F_1$ and $F_2$ which requires running many realizations.
 Moreover, the impact of choosing the $x^{(i)}_j$ (the target BSs in the handover case or choosing beam tracking procedure) propagates in time and can affect the UEs' performance in the next time slots. Therefore, we need to consider the long-term benefits of selecting association indicators besides their immediate effects on the UEs' performance. Furthermore, In order to select the target BSs, we need to model or predict the UEs' performance in the next time slots, which can add more complexity to the network due to the mobility of the UE and obstacles in mmWave networks. These motivate us to utilize the RL to approximate the solution of \eqref{optimization}.


\section{ Proposed Method}\label{BF}
We transform the problem \eqref{optimization} to an RL problem in which the objective function is turned into a reward function, and the constraints are transformed into the feasible state and action spaces.
In the following, first, we start with defining the Markov decision process, and then we will describe our joint handover and beam tracking algorithm.

\subsection{Markov Decision Process Formulation}
RL problems are formulated based on the idea of the Markov decision process (MDP), which is the agent's interaction with different states of the environment to maximize the expected long-term reward. The agent is the main decision-maker who can sit on the edge cloud. All BSs are connected to the agent. Now, we define different elements of an MDP.

\subsubsection{State Space}
The state space describes the environment by which the agent is interacting through different actions. We define the state at time slot $i$ as $s^{(i)}=(\ell^{(i)}), j^{(i)}_S, \text{SNR}^{(i)}, I^{(i)})\in \mathcal{S}$, where $\ell^{(i)}$ is the location index of the UE along the trajectory \footnote{Note that, we discretize the location of the UE along the trajectory. Hence, every location dimension $(x,y)$ a trajectory with length $M$ is mapped to a location index $\ell^{(i)}\in [M]$.}, $j^{(i)}_S$ is the index of the serving BS, $\text{SNR}^{(i)}$ is the SNR value of the UE with serving BS $j^{(i)}_S$ in time slot $i$. $I^{(i)}\in \{0,1\}$ is the beam tracking activation indicator. $I^{(i)}=1$ means the $i$-th time slot is the tracking slot for the UE. 

\subsubsection{Action Space}
The action space includes all possible actions that can be taken by the agent. The action can change the state of the environment from the current state to the target state. In our problem, $a^{(i)}\in \mathcal{A}= \{0,1,2,...,[|\mathcal{B}|]\}$ is the decision regarding beam tracking ($a^{(i)}=0$) or choosing the index of new serving BS in the case of handover decision ($a^{(i)}\in [|\mathcal{B}|]$). In other words, if $a^{(i)}\neq 0$ means the handover decision is made and the value of $a^{(i)}$ shows the target BS. 
Hence, the action is to specify a serving BS for the UE along its trajectory. 

\subsubsection{Policy} A policy $\pi(.)$ maps the state of the environment to the action of the agent. In our case, $\pi$ is a function from $\mathcal{S}$ to $\mathcal{A}$, i.e., 
$ \pi: \mathcal{S} \rightarrow \{0,1,...,[|\mathcal{B}|]\}$

\subsubsection{Rewards} The agent obtains the reward after taking an action $a^{(i)}$ when current state is $s^{(i)}$ and moves to next state $s^{(i+1)}$. Here we define reward $r(s^{(i)}, a^{(i)}, s^{(i+1)})$ as 
\begin{IEEEeqnarray}{rCl}
{r(s^{(i)}, a^{(i)}, s^{(i+1)})} =  \Gamma^{(i)}
- \lambda  1\left\{\Gamma^{(i)}\leq \Gamma_{\rm thr}\right\},
 \label{eqn:rt}
\end{IEEEeqnarray}
where $\Gamma^{(i)}$ is defined in \eqref{throughput}.

\subsubsection{State-action value} The function $Q_{\pi}(s,a)$ is the long-term reward and is defined as the expected summation of discounted reward in the future for the action $a\in \mathcal{A}$  that agent takes in state $s$ under policy $\pi$.
The RL algorithm aims to choose the optimal policy $\pi^\star$ in each state $s$ that maximizes the $Q_{\pi}(s,a)$. With discount factor $\eta \in [0,1]$, we have

\begin{equation*}
Q_{\pi}(s,a)=\mathbb{E}\left\{ \sum_{i} \eta^i r(s^{(i)}, s^{(i)}, s^{(i+1)})\right\},
\end{equation*}
where the expectation is over the transition probabilities.  In our problem, transition probabilities model the SNR variations due to the randomness of the channel fading and blockage. We assume mobility information including the UEs' current location and its trajectory is known\footnote{Note that the location information can be easily fed back through lower-frequency links.}. Therefore, the transition to the next location is deterministic.

The optimal policy in state $s\in \mathcal{S}$ is found by

\begin{equation}\label{optimalpolicy}
\pi^\star(s)=\argmax_{a\in\mathcal{A}} Q_{\pi}(s,a).
\end{equation}
Due to the continuous and large number of state spaces, we apply deep Q-learning (DQL) \cite{Bertsekas} to solve \eqref{optimalpolicy}. In DQL, the state-action value function is estimated by the deep neural network function approximators.

\subsection{Joint Handover and Beam Tracking Algorithm}
Algorithm 1 describes our proposed joint handover and beam tracking algorithm along a trajectory with duration $M$. If the current association cannot offer the required SNR level, the decision regarding handover or beam track is made based on $a^{(i)}$ as the output of the RL algorithm. In the case of the handover decision, the value of $a^{(i)}$ represents the target BS. 

The beam tracking algorithm based on small spatial measurement in time slot $i$ is shown in Algorithm 2. In slot $i$, the algorithm starts by using the main beam of the same serving BS in the previous time slot $i-1$. If the SNR value is lower than the threshold, then starts a small spatial measurement over the AoD direction of the main beam. To quantify the size of the spatial neighbourhood, we define $\Delta \phi$ and $\Delta \theta$ as the maximum absolute horizontal and vertical deviation from the main AoD direction.
We define $\delta \phi$ and $\delta \theta$ as the measurement resolution in horizontal and vertical, respectively. Inspired by \cite{steer}, the spatial neighbourhood $\mathcal{N}$ surrounding the main AoD direction can be expressed using the horizontal neighbourhood $\mathcal{N}_\phi$ and vertical neighbourhood $\mathcal{N}_\theta$ as
\begin{equation}
    \mathcal{N}_\phi (\Delta \phi, \delta \phi)=\left \{i. \delta \phi : i \in \left[ -\left \lfloor  \frac{\Delta \phi}{\delta \phi}\right\rfloor,\left \lfloor \frac{\Delta \phi}{\delta \phi}\right\rfloor \right]\right\}
\end{equation}

\begin{equation}
    \mathcal{N}_\theta (\Delta \theta, \delta \theta)=\left \{j. \delta \theta : j \in \left[ -\left \lfloor  \frac{\Delta \theta}{\delta \theta}\right\rfloor, \left \lfloor \frac{\Delta \theta}{\delta \theta}\right\rfloor \right]\right\}
\end{equation}
where $\lfloor . \rfloor$ is the floor operation. The complete neighbourhood is the Cartesian product of the horizontal and vertical neighbourhoods as

\begin{IEEEeqnarray}{l} \label{eqn:N}
   \mathcal{N}(\Delta \phi, \Delta \theta, \delta \phi, \delta \theta )=\mathcal{N}_\phi (\Delta \phi, \delta \phi) \times \mathcal{N}_\theta (\Delta \theta, \delta \theta)
   \nonumber\\ \qquad \qquad \quad 
   =\{(\phi, \theta) : \phi \in \mathcal{N}_\phi (\Delta \phi, \delta \phi), \theta \in \mathcal{N}_\theta (\Delta \theta, \delta \theta) \}
   \IEEEeqnarraynumspace
\end{IEEEeqnarray}
The spatial neighborhoods $\mathcal{T}^{(i)}$ in time slot $i$ surrounding the main AoD directions  $(\phi^{(i-1)}_{\ell^\star}, \theta^{(i-1)}_{\ell^\star})$ in previous time slot is

\begin{equation}
    \mathcal{T}^{(i)}=(\phi^{(i-1)}_{\ell^\star}, \theta^{(i-1)}_{\ell^\star},)+\mathcal{N}(\Delta \phi, \Delta \theta, \delta \phi, \delta \theta ).
    \label{Tneigh}
\end{equation}
Now given the main AoD direction, we need to find the transmit direction from neighbourhoods $\mathcal{T}^{(i)}$  that offers the SNR threshold. We represent the sorted direction pairs as $[\mathcal{T}^{(i)}]_{\mathcal{I}}$, where $\mathcal{I}$ is the sorted indices. It means the directions in $[\mathcal{T}^{(i)}]_{\mathcal{I}}$ increase in distance from the main AoD direction. Starting from the main AoD direction, the SNR of each transmit direction in  $[\mathcal{T}^{(i)}]_{\mathcal{I}}$ is measured until a beam pair meets the required SNR level. Afterwards, no further measurements are required. If no direction meets the threshold, the entire $(\Delta \phi, \Delta \theta)$-neighbourhood is measured to find the beam pairs that offer the SNR threshold. 

Note that in the worse scenario, if the selected target BS based on our proposed algorithm cannot offer the required SNR level due to very sudden blockage, the conventional handover methods based on searching over the candidate BSs in UEs vicinity can be applied. However, as shown in the numerical results, such extreme case is rare.  
        
        \begin{algorithm}[tp]
	\caption{Joint handover and beam tracking} \label{alg:neighbor}
	\textbf{Input}: Trajectory with duration $M$
	\begin{algorithmic}[1]
	
	\State{Initialization: for $i=1$ set $j^{(1)}_S$=1}
	    \For{$i \in {1,...,M}$}
	   \If{$\text{SNR}^{(i)}_{j_S}< \text{SNR}_{\text{thr}}$}
    	    \State{ Choose the optimal action $a^{(i)}$ based on current $s^{(i)}$.}
    	                \If{$a^{(i)}\neq 0$}. \Comment{handover execution}
    	              \State{  Set $j^{(i)}_S=a^{(i)}$ and run the initial beam training process and compute the achieved throughput $\Gamma^{(i)}$ as \eqref{throughput}}.
    	              \Else    
    	               \State{Run Algorithm 2 and compute $\Gamma^{(i)}$. }
    	  
    	            \EndIf
    	            \EndIf
	   \EndFor
    \end{algorithmic}
    \textbf{Output}: $\Gamma^{(i)}$ 
\end{algorithm}

\begin{algorithm}[tp]
	\caption{Beam tracking in time slot $i$ at the BS $j$} \label{alg:neighbor}
	\textbf{Input}: $[\mathcal{T}^{(i)}]_{\mathcal{I}}$,  $\text{SNR}_{\text{thr}}$, duration of each beam pair testing ($\beta$), $\text{cnt}^{(i)}=0$.
	\begin{algorithmic}[1]
	    \For{$(\phi,\theta) \in [\mathcal{T}]_{\mathcal{I}}$}
	    
		\State{Set $\mathbf{f}_j^{(i)}=\mathbf{a}(\phi,\theta)$.}
		\State{  Measure $\text{SNR}_j^{(i)}$ as \eqref{SNR}.}
	    	\State{  Set $\text{cnt}^{(i)}=\text{cnt}^{(i)}+1$.}  \Comment{number of beam pair testing}
    	                \If{$\text{SNR}^{(i)}_j >=\text{SNR}_{\text{thr}}$}
    	              \State{  $( \phi^{(i)}_{\ell^\star}, \theta^{(i)}_{\ell^\star}
        )=(\phi^\BS,\theta^\BS )$}
        \State{$\tau_b^{(i)}=\beta.\text{cnt}^{(i)}$}
    	                   \State{\textbf{break;}}
    	            \EndIf
    	            \EndFor
    	       
    \end{algorithmic}
    {compute the achieved throughput $\Gamma^{(i)}$ as \eqref{throughput}}
\end{algorithm}

\section{Numerical Results}
\label{results}
\begin{figure}[!t]
    \centering
    \includegraphics[scale=0.18]{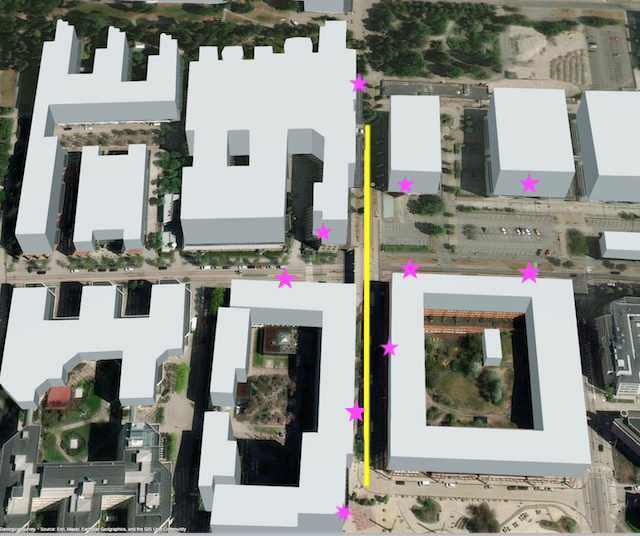}
    \caption{Simulation area in Kista, Stockholm. The yellow line shows the trajectory. Stars show the location of the BSs.}
    \label{fig:scenario}
\end{figure}

We evaluate the performance of the proposed method in an urban environment using the ray tracing tool in the MATLAB toolbox. The output of the ray tracing tool is the $L$ available paths between a BS and a UE in a specific location. 
The ray tracing maintains the spatial consistency of mmWave channels. 
As depicted in Fig. \ref{fig:scenario}, we extracted the building map of Kista in Stockholm city, Sweden and used it as the input data for the ray tracing simulation. In our scenario, we assumed the building material is \emph{brick} and the terrain material is \emph{concrete}. We also add some random obstacles in the street with different heights ($1$ m and $3$ m) and widths ($2$ m and $4$ m) as the human bodies and various vehicles. These temporary obstacles are distributed randomly in the street with density $10^{-2}$ per $m^2$. The material loss and the location of the temporary obstacles are chosen randomly in each realization of the channel. 
The BSs are located on the wall of buildings. The location of the BSs is chosen randomly while covering the entire trajectory.  The BSs' height is \mbox{$6$ m}. We consider a pedestrian mobility model with a speed of $1$ m/s. We consider the different lengths of the trajectories as $100T_{\text{A}}$, $200T_{\text{A}}$, $300T_{\text{A}}$, $400T_{\text{A}}$, $500T_{\text{A}}$. 
The main simulation parameters are listed in Table \ref{table1}.

In the simulation,  we consider the $\text{SNR}_{\text{thr}}=2$ dB and the throughput threshold $\Gamma_{\text{thr}}=1$ bit/Hz. The value of $\tau_c$ is $10$ $ms$. In the case of handover, we fix the initial beam training duration as $\tau_b=\frac{1}{3}\tau_c$. In the case of beam tracking, $\tau_b$ is not fixed and equals the size of measuring neighbourhood multiplied by the duration of each beam pair testing \mbox{($\beta=10$ $\mu s$)}.
We compare the performance of our proposed method with two baselines.
To have a fair comparison, we choose two baselines in which the target BS for the handover is pre-determined. Hence, we do not take into account the discovery time of finding the target BS in the baselines. Just like in our method, the handover is triggered if $\text{SNR}<\text{SNR}_{\text{thr}}$.\newline
As \textbf{Baseline 1} we consider  the multi-connectivity method \cite{multicon3}. We implement a scenario where the UE maintains its connection with a nearby BS as a backup solution while being connected to the serving BS and once it experiences the blockage of the serving link, starts connecting to the backup solution. As \textbf{Baseline 2} we select the learning-based handover in \cite{saraTCCN}. The method shows very good performance in maximizing the achieved rate along the trajectory. In this baseline, the target BS during the handover process is determined by a learning algorithm. Although the target BSs are selected based on the long-term effect on the achieved rate, still can cause frequent handovers and throughput degradation.

First, we fix the number of BSs to $10$ (see Fig. \ref{fig:scenario}). We consider $10^4$ different channel realization as the input of the RL algorithm. After getting the optimal policy, we test it over real-time measurements and report the average of the performance over $500$ channel realizations. Fig. \ref{throubellow} shows the average number of locations with unmet throughput thresholds along the trajectory with different lengths and Fig. \ref{numH} shows the average number of handovers needed. In comparison to the other two baselines, our method provides better throughput results by selecting to perform either beam tracking or a handover. Furthermore, we note that the two baselines have a higher number of handovers than our method due to only considering the handover solution. Hence, by considering the joint handover and beam tracking problem our method  provides better-achieved throughput while decreasing the number of handovers.
\begin{table}[t]
\begin{center}
\caption{Simulation parameters. }
\label{table1}
\begin{tabular}{|c|c|} 
\hline
\textbf{Parameters} & \textbf{Values in Simulations} \\
\hline
\hline
$\BS$ transmit power & 10 dBm\\
Noise power level	&$\sigma^2$=-174 dBm/Hz\\
Signal bandwidth & $100$ MHz \\
BS antenna & $8\times8$ uniform planar array \cite{saraTCCN}\\
Time interval duration & $T_{\text{A}}=1 s$\\
\hline
Neighborhood size & $(\Delta \phi, \Delta \theta)=(10^\circ,10^\circ)$\\
Measurement resolution & $(\delta \phi, \delta \theta)=(5^\circ,5^\circ)$\\
\hline
Discount factor&	$\eta= 0.99$ \\
$\lambda$& 100\\
\hline
\end{tabular}
\end{center}
\end{table}
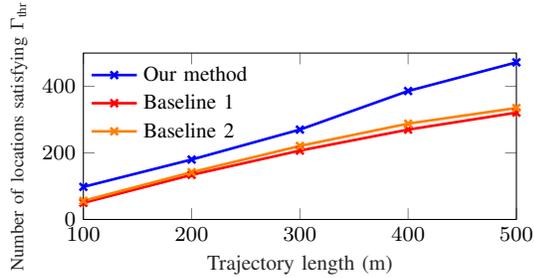
\begin{figure}[t]
	\centering
		{\footnotesize 
%
%
\begin{tikzpicture}

\begin{axis}[%
width=0.65\columnwidth,
height=0.25\columnwidth,
at={(0\columnwidth,0\columnwidth)},
scale only axis,
xmin=100,
xmax=500,
xlabel style={font=\color{white!15!black}},
xlabel={Trajectory length (m)},
ymin=0,
ymax=500,
ylabel style={font=\color{white!15!black}},
ylabel={\scriptsize Number of locations satisfying $\Gamma_{\text{thr}}$},
axis background/.style={fill=white},
legend style={at={(0.41,0.41)}, anchor=south east, legend cell align=left, align=left, draw=white!15!black,draw=none, fill=none}
]
\addplot [color=blue, line width=1.0pt, mark=x, mark size=2.0pt, mark options={solid, blue}]
  table[row sep=crcr]{%
100	98\\
200	180\\
300	270\\
400	386\\
500	472\\
};
\addlegendentry{Our method}

\addplot [color=red, line width=1.0pt, mark=x, mark size=2.0pt, mark options={solid, red}]
  table[row sep=crcr]{%
100	50\\
200	134\\
300	207\\
400	270\\
500	321\\
};
\addlegendentry{Baseline 1}
\addplot [color=orange, line width=1.0pt, mark=x, mark size=2.0pt, mark options={solid, orange}]
  table[row sep=crcr]{%
100	56\\
200	142\\
300	221\\
400	288\\
500	335\\
};
\addlegendentry{Baseline 2}

\end{axis}

\end{tikzpicture}%
}
		\caption{The average number of locations with unmet throughput threshold for different lengths of the trajectory.}
		\label{throubellow}
\end{figure}
\begin{figure}[t]
	\centering
		{\footnotesize 
%
%
\definecolor{mycolor1}{rgb}{0.63529,0.07843,0.18431}%
\begin{tikzpicture}

\begin{axis}[%
width=0.65\columnwidth,
height=0.25\columnwidth,
at={(0\columnwidth,0\columnwidth)},
scale only axis,
xmin=100,
xmax=500,
xlabel style={font=\color{white!15!black}},
xlabel={Trajectory length (m)},
ymin=0,
ymax=10,
ylabel style={font=\color{white!15!black}},
ylabel={\scriptsize Number of handovers},
axis background/.style={fill=white},
legend style={at={(0.41,0.41)}, anchor=south east, legend cell align=left, align=left, draw=white!15!black,draw=none, fill=none}
]
\addplot [color=blue, line width=1.0pt, mark size=2.0pt, mark=x, mark options={solid, blue}]
  table[row sep=crcr]{%
100	0\\
200	0.9\\
300	1.8\\
400	2.5\\
500	4.8\\
};
\addlegendentry{Our method}
\addplot [color=red, line width=1.0pt, mark size=2.0pt, mark=x, mark options={solid, mycolor1}]
  table[row sep=crcr]{%
100	2.6\\
200	3.5\\
300	5.8\\
400	7.5\\
500	8.3\\
};
\addlegendentry{Baseline 1}
\addplot [color=orange, line width=1.0pt, mark size=2.0pt, mark=x, mark options={solid, orange}]
  table[row sep=crcr]{%
100	1.9\\
200 3.1\\
300	4.6\\
400	6.2\\
500	7.6\\
};
\addlegendentry{Baseline 2}
\end{axis}
\end{tikzpicture}%
}
		\caption{The average number of handovers for different lengths of the trajectory.}
		\label{numH}
\end{figure}
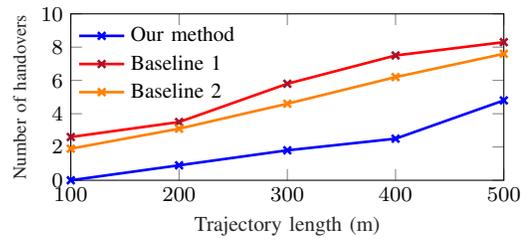
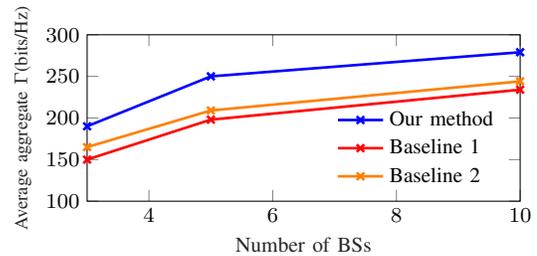
\begin{figure}[t]
	\centering
		{\footnotesize 
%
%
\begin{tikzpicture}

\begin{axis}[%
width=0.65\columnwidth,
height=0.25\columnwidth,
at={(0\columnwidth,0\columnwidth)},
scale only axis,
xmin=3,
xmax=10,
xlabel style={font=\color{white!15!black}},
xlabel={Number of BSs},
ymin=100,
ymax=300,
ylabel style={font=\color{white!15!black}},
ylabel={\scriptsize Average aggregate $\Gamma$(bits/Hz)},
axis background/.style={fill=white},
legend style={at={(0.97,0.03)}, anchor=south east, legend cell align=left, align=left, draw=white!15!black,draw=none, fill=none}
]
\addplot [color=blue, line width=1.0pt, mark=x, mark size=2.0pt, mark options={solid, blue}]
  table[row sep=crcr]{%
3 190	\\
5 250	\\
10 279\\
};
\addlegendentry{Our method}

\addplot [color=red, line width=1.0pt, mark=x, mark size=2.0pt, mark options={solid, red}]
  table[row sep=crcr]{%
3 150	\\
5 198	\\
10 234	\\
};
\addlegendentry{Baseline 1}
\addplot [color=orange, line width=1.0pt, mark=x, mark size=2.0pt, mark options={solid, orange}]
  table[row sep=crcr]{%
3 165	\\
5 209	\\
10 244	\\
};
\addlegendentry{Baseline 2}

\end{axis}

\end{tikzpicture}%
}
		\caption{The average aggregate achieved throughput per Hz along the trajectory with length $300$ m.}
		\label{numBS}
\end{figure}
Fig. \ref{numBS} shows the average aggregate achieved throughput along the trajectory with length $300$ m for different numbers of BSs. By increasing the number of BSs the number of the locations satisfying the $\Gamma_{\text{thr}}$ also increases hence the aggregate throughput along the trajectory increases. Even with a small number of BSs, our method outperforms baselines in aggregate throughput along the trajectory by determining whether to use a handover or beam tracking solution.

We consider 10000 iterations during the training in our method and Baseline 2. With the training machine MacBook Pro 2020 M1 with a memory of 16 GB, each iteration takes about 15 seconds. Note that the absolute value of the training time per iteration depends on the running
machine.

\section{Conclusions} \label{conclusions}
In this work, we proposed and studied a learning-based joint handover and beam tracking method in a mobile mmWave network. The aim of our algorithm is to maximize the aggregate throughput of the UE along a trajectory and ensure the achieved throughput in each location is higher than the threshold. Our evaluation results showed that by making an optimal decision regarding handover execution or beam tracking, our method provides high achievable throughput and reduces the number of handovers. Considering different mobility models and studying the effect of neighbouring size can be valuable future work.

\bibliographystyle{IEEEtran}
\bibliography{IEEEabrv,ref_bib2}
\end{document}